\documentclass[aps,prd,superscriptaddress,eqsecnum,nofootinbib,showkeys,twocolumn,showpacs,preprintnumbers]{revtex4-1}
\usepackage{graphicx}
\usepackage{epsfig}
\usepackage{euscript,amssymb}
\usepackage{amsfonts}
\usepackage{amsmath}
\usepackage{amssymb}
\usepackage{graphicx}
\usepackage{euscript}
\usepackage [usenames, dvipsnames]{xcolor}
\usepackage{cases}
\begin{document}

\title{Slow-Roll Inflation Preceded by a Topological Defect Phase $\bf\grave{a}$ la Chaplygin Gas}

\author{Mariam Bouhmadi-L\'{o}pez}\email{mariam.bouhmadi@ehu.es}
\affiliation{Instituto de Estructura de la Materia, IEM-CSIC, Serrano 121, 28006 Madrid, Spain}
\affiliation{Department of Theoretical Physics, University of the Basque Country
UPV/EHU, P.O. Box 644, 48080 Bilbao, Spain}
\affiliation{IKERBASQUE, Basque Foundation for Science, 48011, Bilbao, Spain}

\author{Pisin Chen}\email{chen@slac.stanford.edu}
\affiliation{Department of Physics \& Graduate Institute of Astrophysics, National Taiwan University, Taipei 10617, Taiwan, R.O.C.}
\affiliation{Leung Center for Cosmology and Particle Astrophysics, National Taiwan University, Taipei 10617, Taiwan, R.O.C.}
\affiliation{Kavli Institute for Particle Astrophysics and Cosmology, SLAC National Accelerator Laboratory, Menlo Park, CA 94025, U.S.A.}
\author{Yu-Chien Huang}\email{r00222014@ntu.edu.tw}
\affiliation{Department of Physics \& Graduate Institute of Physics, National Taiwan University, Taipei 10617, Taiwan, R.O.C.}
\affiliation{Leung Center for Cosmology and Particle Astrophysics, National Taiwan University, Taipei 10617, Taiwan, R.O.C.}
\author{Yu-Hsiang Lin}\email{d00222001@ntu.edu.tw}
\affiliation{Department of Physics \& Graduate Institute of Physics, National Taiwan University, Taipei 10617, Taiwan, R.O.C.}
\affiliation{Leung Center for Cosmology and Particle Astrophysics, National Taiwan University, Taipei 10617, Taiwan, R.O.C.}

\begin{abstract}
We present a simple toy model corresponding to a network of frustrated topological defects of domain walls or cosmic strings that exist previous to the standard slow-roll inflationary era of the universe. Such a network (i) can  produce a slower inflationary era than that of the standard scenario if it corresponds
to a network of frustrated domain walls or (ii) can induce a vanishing universal acceleration; i.e., the universe would expand at a constant speed, if it corresponds to a network of frustrated cosmic strings.
Those features are phenomenologically modeled by a Chaplygin gas that can interpolate between a network of frustrated topological defects and a de Sitter-like or a power-law inflationary era. We show that this scenario can alleviate the quadruple anomaly of the cosmic microwave background spectrum. Using the method of the Bogoliubov coefficients, we obtain the spectrum of the gravitational waves as would be measured today for the whole range of frequencies. We comment on the possible detection of this spectrum by the planned detectors like BBO and DECIGO.
\end{abstract}


\date{\today}
\maketitle

\section{Introduction}
\label{intro}

The inflationary scenario explains several cosmological problems such as the spatial flatness and homogeneity of the universe, and it predicts an almost scale-invariant power spectrum of the primordial density perturbation, which agrees with most of the WMAP observation on cosmic microwave background (CMB) anisotropy. On the other hand, there remains some discrepancies between the standard cosmological model predictions and the WMAP observations on large angular scales. One of them is the suppression of the $l=2$ quadruple mode of the CMB temperature power spectrum (c.f. for example \cite{CMBanomalies}). There have been different approaches to solve this problem. In Refs \cite{Contaldi:2003zv,Boyanovsky:2006pm}, it is considered that the slow-roll inflation was preceded by a fast-roll stage where the kinetic energy of the scalar field was dominant, which leads to a suppression of the quadrupole moment. Such a fast-roll stage can be reached by tuning the initial condition of the inflation, for example. Another way to reach an early stage of kinetic energy dominance was considered in Refs.~\cite{Piao:2003zm,Piao:2005ag} through bounces or cyclic universes where just before the bounce or the new cycle the kinetic energy density of the field dominates over its potential.
Other authors have considered different pre-inflationary phases of the universe to suppress the primordial power spectrum at large scales. For example in Refs.~\cite{Powell:2006yg,  Wang:2007ws}, a radiation-dominated pre-inflationary era was considered and in \cite{Scardigli:2010gm}, a pre-inflation matter era provided by primordial micro black hole remnants was introduced.

The CMB quadruple anomaly affects the lowest modes of the primordial power spectrum; i.e.~ modes with comoving wave numbers of the order of $10^{-3} \, \textrm{Mpc}^{-1}$. Therefore, these are the first modes to exit the horizon and the last ones to reenter. Consequently we expect that these modes are heavily affected by the physics of the very early universe, possibly the physics prior to the slow-roll inflationary era.
Based on this reasoning, we propose a new cosmological period just before the slow-roll inflationary era. This period corresponds to an era described by a network of topological defects, which we will assume to be frustrated domain walls or cosmic strings \cite{Bucher:1998mh, Battye:1999eq,Spergel:1996ai,BouhmadiLopez:2002mc,BouhmadiLopez:2002nz}, for the sake of simplicity. We expect that the production of topological defects at those scales follows the prediction of high energy  physics. In addition, given that the topological defects era proceeds the slow-roll inflationary era, the topological networks will affect mainly the lower modes of the power spectrum of the scalar and tensorial perturbations. Afterwards, they will soon be diluted during most of the inflationary era.

There is a simple way to model the ideas previously mentioned. We consider the matter content of the early universe to be described by a kind of generalized Chaplygin gas (GCG) \cite{Kamenshchik:2001cp, Bilic:2001cg, Bento:2002ps, BouhmadiLopez:2004mp, Chimento:2005au}, which offers a smooth transition between different eras:
\begin{itemize}
\item A period dominated by a network of frustrated domain walls (NFDW) at earlier time and a de Sitter-like phase at later time.
\item A period dominated by a network of frustrated cosmic strings (NFCS)  at earlier time and a de Sitter-like phase at later time.
\end{itemize}
 The idea of an early Chaplygin gas phase in the universe was first suggested in Ref.~\cite{GCGquantum} (see also \cite{Bertolami:2006zg, BouhmadiLopez:2009pu}) and later extended in Refs.~\cite{BouhmadiLopez:2009hv,BouhmadiLopez:2011kw,BouhmadiLopez:2012qp}. The energy density of a network of frustrated topological defects (NFTD) can be described in a compact way, for example, as
\begin{equation}
\rho= \left(\frac{B_1}{a^{\beta _1\left(1+\alpha _1\right)}}+A_1\right)^{1\left/\left(1+\alpha _1\right)\right.},
\label{1.1}
\end{equation}
where $a$ is the scale factor, $B_1$ and $A_1$ are constants related to the energy scale of the NFTD and the de Sitter-like inflationary era, respectively, $\alpha_1$ and $\beta_1$ are constants such that $\beta_1=1,2$ for the network of domain walls and cosmic strings, respectively. We assume that $1+\alpha_1$ is positive such that the inflationary era is preceded by a topological dominance phase. Let us be reminded in this regard that the energy density of NFDW and NFCS scales as $1/a$ and $1/a^2$, respectively \cite{topo}. It is worthy to stress that the NFTD epoch preceding the slow-roll inflationary can in principle produce inflation as well; indeed this is the case for NFDW, but this inflation is much slower, i.e.~much \textit{lazier} than the slow-roll inflation. Moreover, for a NFCS dominated period the universe is increasing its size at a constant speed; i.e.~with no acceleration or deceleration. From now on whenever we refer to a pre-inflationary era, we will be referring to a pre-slow-roll inflationary era.

This paper is organized as follows: In Sec. \ref{model}, in order to solve the CMB quadrupole problem, we propose a NFTD era just before the standard inflationary era. We model this idea using the framework of matter $\rm\grave{a}$ la Chaplygin gas during the early universe as a way of interpolating between a NFTD and a slow-roll inflationary period. We then explain how to constrain our model observationally. In Sec. \ref{sp}, we briefly review the standard cosmological perturbation formalism and obtain numerically the curvature power spectrum as well as the CMB temperature anisotropy spectrum for the model presented in Sec. \ref{model}. In Sec. \ref{tp}, we obtain numerically the spectrum of the gravitational waves for the model  introduced in Sec. \ref{model} by using the method of the Bogoliubov coefficient \cite{Moorhouse:1994nc}. Finally we summarize and conclude in Sec. \ref{sum}.

\section{Model building and Parameters fixing}
\label{model}
We consider a spatially flat Friedmann-Lema\^{i}tre-Robertson-Walker (FLRW) universe filled with the matter content described by Eq.~\eqref{1.1}. The energy conservation gives
\begin{equation}
\dot{\rho}+3H(\rho+p)=0,
\label{eq:EnergyConservation}
\end{equation}
where a dot corresponds to a derivative with respect to the cosmic time and $H$ stands for the Hubble rate. By inserting Eq.~\eqref{1.1} into Eq.~\eqref{eq:EnergyConservation}, one obtains the pressure of the matter content
\begin{equation}
p=\left(\frac{\beta_1}{3}-1\right)\rho -\frac{\beta_1}{3}\frac{A_1}{\rho ^{\alpha_1}}.
\end{equation}

In the Planck unit ($c = \hbar = G = 1$), the Friedmann equation reads
\begin{equation}
 H^2=\frac{\kappa^2}{3}\rho,
\label{fe}
\end{equation}
where $\kappa^2\equiv 8\pi G=8\pi$. The conformal time $\tau$ can be expressed as
\begin{equation}
\tau =\frac{\sqrt{3}}{\kappa }\frac{b}{c}A_1^{-(b+c)}\left(\frac{B_1}{A_1}\right)^by^c_2F_1(c,1-b,c+1,y),
\label{tau}
\end{equation}
where $y=\left[1+\left(B_1/A_1\right)a^{-\beta _1\left(1+\alpha _1\right)}\right]^{-1}$, $b=1\left/\left[\beta _1\left(1+\alpha _1\right)\right]\right.$, $c=(-2+\beta_1)/[2 (1+\alpha_1) \beta_1]$, and $_2F_1$ is a hypergeometric function \cite{W}.  From Eq.~\eqref{tau} we can see that the universe began in NFTD dominated era in the past infinity, and turned into a de Sitter-like space, in which $a \propto 1 / \tau$, at later time.

In order to obtain the power spectrum of the gravitational wave as would be measured today, which will be obtained in section \ref{tp}, we divide the expansion of {\color{black} the} universe into three successive periods: the pre-inflation NFTD dominated era, the inflating phase, and the standard $\Lambda$CDM epoch. The energy density of each of these periods can be modeled as
\begin{subnumcases}{\label{1m} \rho =}
	\left(
		\frac{B_1}{a^{\beta _1 \left( 1 + \alpha_1 \right)}} + A_1
	\right)^%
		{1 \left/ \left( 1 + \alpha_1 \right) \right.},
	\label{1m1} \\
	\left(
		A_2 + \frac{B_2}{a^{4 \left( 1 + \alpha_2 \right)}}
	\right)^%
		{1 \left/ \left( 1 + \alpha_2 \right ) \right.},
	\label{1m2} \\
	\rho_{\textrm{r0}} \left( \frac{a_0}{a} \right)^4
		+ \rho_{\textrm{m0}} \left( \frac{a_0}{a} \right)^3
		+ \rho_{\Lambda}.
	\label{1m3}
\end{subnumcases}
Expression (\ref{1m1}) describes the energy density of the NFTD era, which was introduced in Eq.~(\ref{1.1}), followed by the de Sitter-like inflating phase. The model described by Eq. (\ref{1m2}) was previously studied within an inflationary framework in Ref.~\cite{BouhmadiLopez:2009hv} (see also Ref.~\cite{Chimento:2005au}) and, under suitable constraints on $A_2$, $B_2$, and $\alpha_2$, can depict the transition from the de Sitter-like era to the radiation dominated era. The energy density (\ref{1m3}) is the standard $\Lambda$CDM model, in which $\rho_{\text{r0}}$, $\rho_{\text{m0}}$, and $\rho_\Lambda$ are the energy densities of the radiation, matter, and dark energy today, respectively. As we will show later, the parameters of the model can be constrained using observational data corresponding to the present energy density of radiation, the scalar power spectrum, and the spectral index at a given pivot scale. In addition, by requiring that the energy density is continuous at each transition, we have the conditions
\begin{align}
A_1 &= A_2^{(1+\alpha_1)/(1+\alpha_2)},\\
B_2 &= \left(\rho_{r0}a_0^4\right)^{1+\alpha_2}\label{B2}.
\end{align}

In order to obtain the scalar power spectrum, it is useful to model the matter content in the first two periods of Eq.~(\ref{1m}); i.e. those described by Eqs.~(\ref{1m1}) and (\ref{1m2}), through a scalar field, with the condition that the energy density and pressure of the scalar field are the same as that given by Eqs.~(\ref{1m1}) and (\ref{1m2}). The energy density and pressure of the scalar field are
\begin{equation}
\rho_\phi=\frac{\phi'^2}{2\,a^2}+V(\phi), \qquad p_\phi=\frac{\phi'^2}{2\,a^2}-V(\phi),
\label{rhop}
\end{equation}
where the primes denote the derivatives with respect to the conformal time. The scalar field and its potential in the first period is given by
%
\begin{align}
	\label{phi1} \phi (a) = &\frac{1}{\kappa \left(1+\alpha _1\right)\sqrt{\beta _1}} \notag \\
	&\times \ln \left(\frac{-1+\sqrt{1+(\frac{a}{B_1/A_1})^{\beta _1\left(1+\alpha _1\right)}}}{1+\sqrt{1+(\frac{a}{B_1/A_1})^{\beta _1\left(1+\alpha _1\right)}}}\right), \\
	\label{v1e} V_1(a) = &\frac{V_0}{6}\left\{\left(6-\beta _1\right)\left[1+\left(\frac{B_1/A_1}{a^{\beta _1}}\right)^{\left(1+\alpha _1\right)}\right]^{1\left/\left(1+\alpha _1\right)\right.}\right. \notag \\
	&\left.+\beta _1\left[1+\left(\frac{B_1/A_1}{a^{\beta _1}}\right)^{\left(1+\alpha _1\right)}\right]^{-\alpha _1/\left(1+\alpha _1\right)}\right\},
\end{align}
%
where $V_0=A_1^{1/(1+\alpha_1)}$. Similarly, the scalar field and its potential in the second period can be obtained by replacing $\alpha_1$ with $\alpha_2$ and setting Eq.~(\ref{phi1}) and Eq.~(\ref{v1e}), which gives
%
\begin{align}
	\label{phi2} \phi(a) = &\frac{1}{\kappa \left(1+\alpha _2\right)\sqrt{4}} \notag \\
	&\times \ln \left(\frac{-1+\sqrt{1+(\frac{a}{B_2/A_2})^{4\left(1+\alpha _2\right)}}}{1+\sqrt{1+(\frac{a}{B_2/A_2})^{4(1+\alpha _2)}}}\right),\\
	\label{v2e} V_2(a) = &\frac{V_0}{3}\left\{\left[1+\left(\frac{B_2/A_2}{a^4}\right)^{\left(1+\alpha _2\right)}\right]{}^{1\left/\left(1+\alpha _2\right)\right.}\right. \notag \\
	&\left.+2\left[1+\left(\frac{B_2/A_2}{a^4}\right)^{\left(1+\alpha _2\right)}\right]^{-\alpha _2/\left(1+\alpha _2\right)}\right\}.
\end{align}
%
We can obtain the potential of the scalar field for the two periods as functions of the scalar field by substituting the inverse function of Eq.~(\ref{phi1}) into Eq.~(\ref{v1e}) and similarly {\color{black}that of} Eq.~(\ref{phi2}) into Eq.~(\ref{v2e}), respectively, which leads to 

\begin{eqnarray}
\label{v1}
V_1(\phi )&=&\frac{V_0}{6}\left\{\left(6-\beta _1\right)\cosh \left[\kappa \left(1+\alpha _1\right)\sqrt{\beta _1}\frac{\phi }{2}\right]^\frac{2}{1+\alpha_1}\right. \nonumber \\ &\;&\left.+\;\beta _1\cosh \left[\kappa \left(1+\alpha _1\right)\sqrt{\beta _1}\frac{\phi }{2}\right]^\frac{-2\alpha _1}{1+\alpha _1}\right\},\\
V_2(\phi)&=&\frac{V_0}{3}\left\{\cosh  \left[\kappa \left(1+\alpha _2\right)\phi \right]^{\frac{2}{1+\alpha _2}} \right. \nonumber \\ &\;&\left. +2\cosh \left[\kappa \left(1+\alpha _2\right)\phi \right]^{\frac{-2\alpha _2}{1+\alpha _2}}\right\}.
\label{v2}
\end{eqnarray}
Eq.~(\ref{v2}) coincides with the potential in Ref.~\cite{{BouhmadiLopez:2009hv}}, as it should be. We connect Eq.(\ref{v1}) and Eq.(\ref{v2}) at $\phi=0$ so that the potential and its first derivative with respect to $\phi$ {\color{black}are analytically} continuous at the connecting point. The result is shown in Fig.~\ref{v_v0}.

\begin{figure}[t]
\centering
\includegraphics[width=8cm]{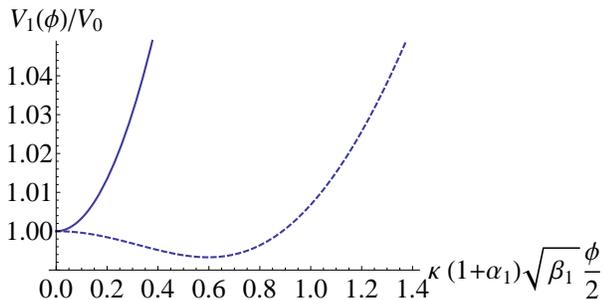}
\centering
\caption{The scalar field potential for the NFDW era for different values of $\alpha_1$. The solid curve corresponds to $\alpha_1$=1 and the dashed curve corresponds to $\alpha_1$=7. It is similar for the case of NFCS, which we omit for simplicity.}
\label{alpha}
\end{figure}

\begin{figure}[t]
\centering
\includegraphics[width=8 cm]{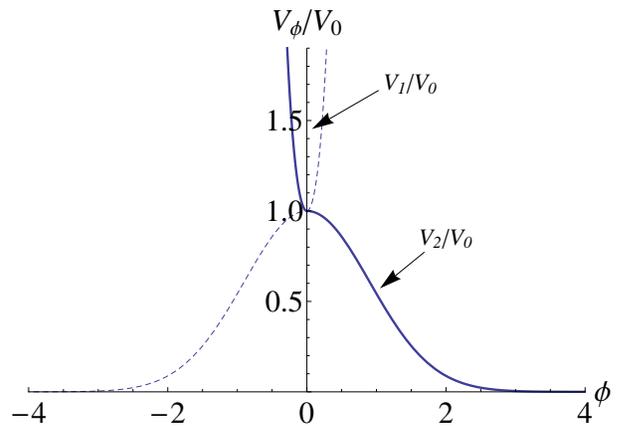}
\centering
\caption{The scalar field potential for $\beta_1=1$. The upward concave curve is $V_1$ (cf.~Eq.~(\ref{v1})) and the downward convex curve is $V_2$ (cf.~Eq.~(\ref{v2})). We can choose $\phi\leq0$ for $V_1$ and $\phi\geq0$ for $V_2$ to describe the potential of the scalar field $\phi$.}
\label{v_v0}
\end{figure}

Next we tackle the issue of analyzing the potentials (\ref{v1}) and (\ref{v2}).
First of all, we consider that the scalar field potential (\ref{v1}) has a unique minimum at $\phi=0$ to maximize the amount of inflation during the first period. Notice that unless this condition is imposed, the scalar field might roll down the potential till it reaches the minimum of $V_1(\phi)$ and then would have to climb up to reach the local maximum located at $\phi=0$, as shown in Fig.~\ref{alpha}. Imposing that the potential (\ref{v1}) has a unique minimum reached at $\phi=0$ implies a condition on the parameters $\alpha_1$ and $\beta_1$ that
\begin{equation}
\alpha_1 < \frac{6-\beta_1}{\beta_1}.
\label{conditionalpha_1}
\end{equation}
Therefore, bearing in mind that (i) $\beta_1=1,2$ for NFDW and NFCS, respectively, and (ii)  $0<1+\alpha_1$ so that the phase of FNTD precedes the inflationary phase, we conclude that $-1<\alpha_1<5$ for NFDW and $-1<\alpha_1<2$ for NFCS. We show the shape of the potential $V_1(\phi)$ for different cases when the condition (\ref{conditionalpha_1}) is fulfilled and violated in Fig.~\ref
{alpha}.


In addition, we can constrain our model; i.e. the potentials (\ref{v1}) and (\ref{v2}), using the methodology used by one of us in Ref.~\cite{BouhmadiLopez:2009hv}. More precisely,  we can use the WMAP7 observation of the power spectrum of the comoving curvature perturbation, $P_s=2.45\times10^{-9}$, and the spectral index, $n_s=0.963$, at the pivot scale $k_0=0.002$ Mpc$^{-1}$ to fix the parameters in our model \cite{Komatsu:2010fb}. We can as well impose a bound on the number of e-folds, $N_c$, since a given mode exits the horizon until the end of inflation as done in Ref.~\cite{BouhmadiLopez:2009hv}.
This gives the best-fit values for $\alpha_2$, $V_0$, and, therefore, $A_2$. Notice that once $V_0$ is fixed, the parameter $A_1$ is fixed for a given $\alpha_1$ as well, since $V_0=A_1^{1/(1+\alpha_1)}$.

The parameter $B_1$ in Eq.~(\ref{1m1}) fixes the energy density of the NFTD, which strongly affects the lowest modes that exited the horizon around the onset of inflation, and causes a significant drop on the lowest modes of the primordial spectrum of the curvature perturbation. Although we expect that the NFTD would affect the lowest modes, we must make sure that the observed  curvature power spectrum $P_s$ and the spectral index $n_s$ at the pivot scale $k_0$ are recovered.
Therefore, we choose the value of $B_1$ such that $P_s$ and $n_s$ correspond to the correct values at $k_0$, and the amplitude of $P_s$ drops at the scales whose comoving wave numbers are smaller than $k_0$. Roughly speaking, the parameter $B_1$ controls the horizontal shift of the curvature power  spectrum.

However, following this procedure, it turns out that we can not find a set of values for the parameter $B_1$ that satisfies the constraint on the spectral index. In fact, in this model $B_1$ turns out to be always smaller than $0.9$. The reason of this drawback is the second period described by Eq.~(\ref{1m2}), which corresponds to a short transition from a de Sitter era to a radiation dominated period. More precisely, the model described by Eq.~(\ref{1m}) does not give enough e-folds during the inflationary era. We show, as an example, in Fig.~\ref{v_v0} how the scalar field rolls too quickly and the radiation dominated phase is reached too early in the case corresponding to a NFDW.

We therefore suggest an alternative model described by
\begin{subnumcases}{\label{2m}\rho =}
 \frac{B_1}{a^{\beta_1}}+\left(\frac{A_2}{a^{(1+\beta_2 )}}\right)^{\frac{1}{1+\beta_3 }}\label{2m1}, \\
\left(\frac{A_2}{a^{1+\beta_2 }}+\frac{B_2}{a^{4(1+\beta_3 )}}\right)^{\frac{1}{1+\beta_3 }}
\label{2m2}, \\
 \rho _{\text{r0}}\left(\frac{a_0}{a}\right){}^4+\rho _{\text{m0}}\left(\frac{a_0}{a}\right){}^3+\rho _{\Lambda }
\label{2m3},
\end{subnumcases}
where $\beta_1=1,2$ discriminates the NFDW and the NFCS as in the former model, and $\beta_2, \beta_3, B_1, A_2, B_2$ are constants we will explain below. We choose this model for the second period since it generates an almost flat curvature spectrum for modes larger than the pivot scale $k_0=0.002$ Mpc$^{-1}$ and gives enough e-folds during the power law inflationary period. In addition, the model introduces in a natural way that a NFTD precedes the power law inflation as $(1+\beta_2)/(1+\beta_3)<\beta_1$ (please see also the conditions (\ref{c1}), (\ref{c2}) and (\ref{c3})).

The first period in Eq.~(\ref{2m1})  describes the matter content of the universe during a period that transits from a NFTD dominated phase to a power-law inflationary era. The parameters $B_1$ and $A_2$ are associated with the energy scale of the NFTD and the power-law inflation, respectively.  The second period with the energy density (\ref{2m2}) was previously studied within another inflationary framework in Ref.~\cite{BouhmadiLopez:2011kw}. It connects smoothly a power-law inflating phase with a radiation dominated universe, and the constraints on the parameters $\beta_2$ and $\beta_3$ are,\footnote{The notation is different from the one used in the work \cite{BouhmadiLopez:2011kw}. The parameters $\beta$ and $\alpha$ in Ref.~\cite{BouhmadiLopez:2011kw} are denoted as $\beta_2$ and $\beta_3$ here, respectively.}
\begin{eqnarray}
\label{c1} 1+\beta_2&<&0, \\ 
\label{c2} 1+\beta_3&<&0, \\ 
\label{c3} 2(1+\beta_3)&<&1+\beta_2 .
\end{eqnarray}
These constraints imply that (i) there is a power-law inflating phase, (ii) the inflationary era precedes the radiation dominated period, and (iii) the null energy condition is always fulfilled so that there is no super-inflationary phase. Finally, the energy density  described by (\ref{2m3}) corresponds to the $\Lambda$CDM model as that described by Eq.~(\ref{1m3}).

Although there seems to be many free parameters, they can be fixed down to only one by the following procedure: (i) Fix $B_2$ by the current amount of radiation for a given $\beta_3$. (ii) Constrain the power-law expansion quantified by $A_2^{(1+\beta_3)/(1+\beta_2)}$ by the WMAP7 data of the curvature power spectrum $P_s$ and the spectral index $n_s$. Since there are three parameters ($A_2$, $\beta_2$ and $\beta_3$) to be constrained by only two conditions ($P_s$ and $n_s$), we are left with the only one free parameter which we choose to be $\beta_3$ for convenience. (iii) Fix $B_1$ such that $P_s$ and $n_s$ give the correct value at the pivot scale $k_0$ and the drop on $P_s$ just begins at comoving wave numbers smaller than $k_0$.

Again, it is suitable to introduce a scalar field that mimics the matter content described in Eqs.~(\ref{2m1}) and (\ref{2m2}); i.e.
we describe the dynamics of the model through a scalar field with a potential whose energy density and pressure can be obtained from Eq.~(\ref{rhop}).  During the NFTD period (cf. Eq.~(\ref{2m1})),  the mapping between the scalar field $\phi$ and the perfect fluid of our model leads to
\begin{eqnarray}
\label{phia1}
\phi(a)&=&\sqrt{v}\coth ^{-1}\sqrt{\frac{\beta _1}{v }} \nonumber\\
&\,&-\sqrt{\beta _1}\tanh ^{-1}\sqrt{\zeta+(1-\zeta)\frac{v }{\beta _1}}, \\
\label{v21}
V_1(a)&=&\frac{1}{6} \left[\left(5+\frac{ \beta_3 -\beta_2 }{1+\beta_3 }\right)\left(\frac{A_2}{a^{1+\beta_2 }}\right)^{\frac{1}{1+\beta_3 }} \right. \nonumber\\
&\,\,& \left.+\, (6-\beta_1)\frac{B_1}{a^{\beta_1}}\right],
\end{eqnarray}
where $\zeta=[1+(A_2^{1\left/\left(1+\beta _3\right)\right.}/B_1)a^{\beta _1-v }]^{-1}$, $v=(1+\beta_2 )/(1+\beta_3 )$, and $V_1(\phi)$ stands for the scalar field potential during this period.
Similarly, we map the perfect fluid with the energy density (cf. Eq.~(\ref{2m2})) to the scalar field $\phi$ with  a new potential $V_2(\phi)$  \cite{BouhmadiLopez:2011kw}
\begin{eqnarray}
\phi(a)&=&\frac{1}{q \kappa }\left[4 \tanh^{-1}\sqrt{1+\frac{q}{4(1+\beta_3 )}\frac{1}{1+\xi }}-2  \right. \nonumber\\ &\,&\left.
 \sqrt{\zeta}\coth^{-1}\sqrt{\frac{4}{\zeta}\left(1+\frac{q}{4(1+\beta_3 )}\frac{1}{1+\xi }\right)}\right], \nonumber\\
\label{phia2}\\ \nonumber
\label{v22}
V_2(a)&=&A_2^{1/(1+\beta_3 )}\left(\frac{A_2}{B_2} \right)^{-\zeta /q}(1+\xi )^{1/(1+\beta_3 )}
\\ \nonumber &\,&\xi ^{-\zeta/q} \left(\frac{1}{3}-\frac{q}{6(1+\beta_3 )}\frac{1}{1+\xi }\right), \nonumber\\
\label{v22}
\end{eqnarray}
where $\xi =(B_2/A_2)a^q$ and $q=1+\beta_2 -4(1+\beta_3 )$. The potential (\ref{v22}) was previously obtained in Ref.~\cite{BouhmadiLopez:2011kw}. Such a potential, with an appropriate initial condition, drives a power law inflation and mimics a radiation dominated universe afterwards.

\begin{figure}[t]
\centering
\includegraphics[width=8cm]{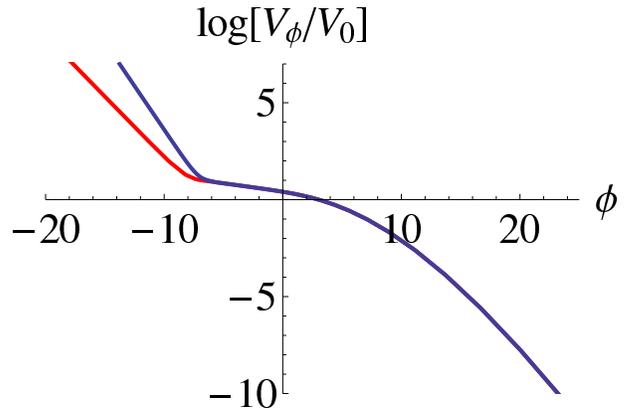}
\centering
\caption{This plot shows the rescaled potentials given in Eqs.~(\ref{v21}) and (\ref{v22}) versus the scalar field $\phi$. The blue curve corresponds to the NFCS case. The red curve corresponds to the  NFDW case and  $V_0=A_2^{1/(1+\beta_3 )}\left(A_2/B_2 \right)^{-(1+\beta_2 )/(q(1+\beta_3 ))}$. The energy scale of inflation, $V_0$, in our model is about $10^{15}$ GeV for both NFDW and NFCS.}
\label{potential}
\end{figure}

Unlike the previous model described by Eq.~(\ref{phi1})-(\ref{v1e}) and Eq.~(\ref{phi2})-(\ref{v2e}), here it is not feasible to find analytically the inverse functions of Eq.~(\ref{phia1}) and Eq.~(\ref{phia2}), so we cannot obtain the analytical forms of the potential as functions of $\phi$. {\color{black}We thus} connect the scalar field potential numerically. $V_1(a)$ and $V_2(a)$ are connected at the intersection of the first two periods (Eq.~(\ref{2m1}) and Eq.~(\ref{2m2})), {\color{black}where} the second term of Eq.~(\ref{2m1}) dominates over {\color{black}its first term, and the first term of Eq.~(\ref{2m2}) dominates over its second term} so that the potential and its first derivatives with respect to $\phi$ are approximately continuous at the intersection of the first two periods. It is worthy to notice that an integration constant appears when we integrate Eq.~(\ref{rhop}) after mapping it to the energy density and pressure of a given perfect fluid. Therefore, we can always choose the constant properly such that the scalar field $\phi$ is continuous at the connecting point. As a result, we can use the scale factor $a$ as a parametric parameter to plot $V_1(\phi)$ and $V_2(\phi)$, which {\color{black}are} shown in Fig.~\ref{potential} as an example. The scalar field starts with a negative value and rolls down the potential as the universe inflates until it reaches the radiation dominated era.

\section{scalar perturbations}
\label{sp}
The accelerated expansion of the universe during the primordial inflationary  era converts the initial quantum fluctuations in the universe into macroscopic cosmological perturbations, which leads to the inhomogeneity we observe nowadays in the CMB \cite{Langlois:2010xc,Mukhanov:1990me}.
Following the standard approach, we will use gauge invariant quantities that involve the metric perturbations and the scalar field fluctuations \cite{LiddleLyth}. For convenience, we will choose the comoving curvature perturbation, $R$, which in addition is conserved on large scales \cite{Wands:2000dp, Lyth:2003im}.

We expect that a NFTD in the very early universe and just before the inflationary era could give the appropriate corrections to the quadrupole modes of the CMB data as observed nowadays \cite{Komatsu:2010fb}. We will next quantify the quantum cosmological perturbations during that period and obtain the power spectrum of the scalar perturbations.
\subsection{Formalism}
The scalar perturbations can be described by introducing the variable (see, for example, Ref.~\cite{Bassett:2005xm})
\begin{equation}
 u=z R,
\end{equation}
where $z\equiv \frac{a\dot{\phi}}{H}$. The variable $u$ can be decomposed into Fourier modes, $u_k$, which fulfill the field equation \cite{Bassett:2005xm}
\begin{equation}
\frac{d^2u_k}{d\tau ^2}+\left(k^2-\frac{1}{z}\frac{d^2z}{d\tau ^2}\right)u_k=0.
\label{waveeq}
\end{equation}
The modes $u_k$ can be mapped to the spectrum of the comoving curvature perturbations which reads \cite{Bassett:2005xm}
\begin{equation}
 \frac{2\pi ^2}{k^3}P_R(k)=\frac{\left|u_k\right|^2}{z^2}.
\end{equation}
Given that we are dealing with adiabatic perturbations, the comoving curvature perturbations remain constant on large scales and consequently we can equate the power spectrum at the horizon exit with the power spectrum of the primordial scalar perturbations at the horizon reentry  as observed on the CMB. Therefore, for a given mode $k$, the spectrum is evaluated at the horizon exit; i.e. when $k=a_{\text{cross}} H$, where $a_{\text{cross}}$ stands for the value of the scale factor when the mode exists the horizon.

\subsection{Numerical Solutions}
We next obtain the evolution of the mode function $u_k(\tau)$ for each comoving wave number $k$ in order to obtain the curvature perturbation spectrum. We will tackle this issue numerically rather than using the standard results for slow-roll inflation \cite{Langlois:2010xc,Lidsey:1995np,Bassett:2005xm}, because those conditions are not fulfilled at very early time when the NFTD is dominant.
It is easier to solve  Eq.~(\ref{waveeq}) numerically by splitting it into two first order differential equations,
\begin{eqnarray}
\begin{cases}
 X^\prime=Y\\
 Y'=-\left(k^2-\frac{z^{\prime\prime}}{z}\right)X,
\end{cases}
\label{wavesf}
\end{eqnarray}
where we have set $X=u_k$.

\begin{figure}[t]
\centering
\includegraphics[width=8cm]{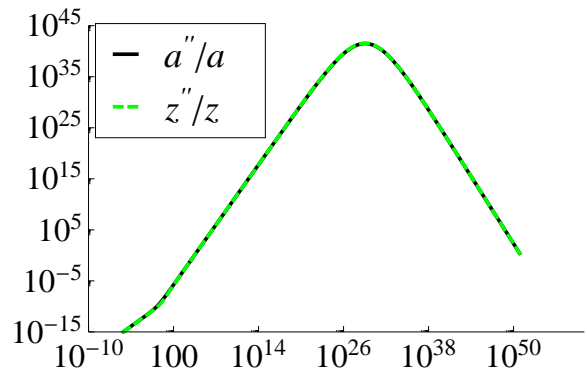}
\centering
\caption{The green dashed curve corresponds to $z^{\prime\prime}/z$ and the black curve corresponds to $a^{\prime\prime}/a.$  As can be seen that the approximation  $z^{\prime\prime}/z \approx a^{\prime\prime}/a$ holds during the NFDW dominance and the power law inflationary era.}
\label{dwzpp_z}
\end{figure}
\begin{figure}[t]
\centering
\includegraphics[width=8cm]{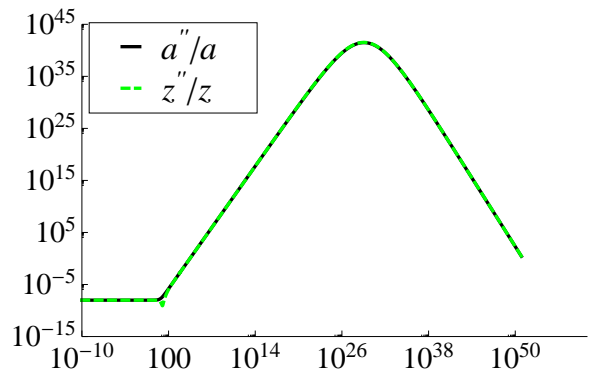}
\centering
\caption{The green dashed curve corresponds to $z^{\prime\prime}/z$ and the black curve corresponds to $a^{\prime\prime}/a.$ As can be noticed that the approximation  $z^{\prime\prime}/z \approx a^{\prime\prime}/a$ holds during the NFCS dominance and the power law inflationary era.}
\label{cszpp_z}
\end{figure}

In addition, we need to impose a set of boundary conditions at the time when the wavelength of a given mode $k$ is much smaller than the Hubble radius; that is, $k\gg a H$. Those boundary conditions will depend on the specific NFTD scenario and also on the given scale of the mode, as we will explain shortly.
It is also worthy to stress two things: (i)  the approximation  $z^{\prime\prime}/z \approx a^{\prime\prime}/a$ holds during the NFTD dominance and during the power law inflationary era
(See Fig.~\ref{dwzpp_z} and Fig.~\ref{cszpp_z}), and (ii) the analytical solutions for the power-law inflation \cite{Sa:2008yq} can be used as boundary conditions for the  modes with comoving wave numbers larger than roughly 10$^{-3}$ Mpc$^{-1}$ as explained next. We first remind that the power law solutions for $X$ and $Y$ read \cite{Lidsey:1995np}
\begin{align}
X(\tau)&=\frac{\sqrt{-\pi \tau }}{2}H_{\frac{1}{2}-l}^{(1)}(-k\tau ), \label{pw1}\\
Y(\tau)&=\frac{-\sqrt{-\pi  \tau}k}{2} H_{\frac{1}{2}-l }^{(1)}(-k\tau )+\frac{l}{2}\sqrt{\frac{-\pi }{\tau}}H_{\frac{1}{2}-l }^{(1)}(-k\tau ), \label{pw2}
\end{align}
where $l$ is the exponent characterizing the power law expansion in terms of the conformal time $\tau$; i.e. $a \propto \tau^l$. For our model, $l=\left.((1+\beta_2)/(2(1+\beta_3))-1\right.)^{-1}$. For those modes whose $k\geq10^{-3} \, $Mpc$^{-1}$,  we can start the numerical integration of Eq.~(\ref{wavesf}) during the power-law inflationary dominated era where the condition $k\gg a H$ is still satisfied. However, for scales roughly smaller  than 10$^{-3}$ Mpc$^{-1}$, we split the boundary conditions imposed on the NFDW and the NFCS separately. Despite the general solutions (\ref{pw1}) and (\ref{pw2}) are still valid for the NFTD, the exponent describing the power law expansion, $l$, depends on the specific characters of the NFTD:
\begin{itemize}
\item During NFDW dominant era, $\rho\sim B_1/a^{\beta_1}$ with $\beta_1=1$, which  also implies a power-law expansion because $a(\tau)\propto \tau^{1 / \left( \beta_1/2-1 \right)}$. Thus we can use the solutions (\ref{pw1}) and (\ref{pw2}) as boundary conditions with $l=1/(\beta_1/2-1)$.

\item During the NFCS dominant era,  $\rho\sim B_1/a^{\beta_1}$ with $\beta_1=2$. Notice that for this value of $\beta_1$ the exponent $l$ in Eqs.~(\ref{pw1}) and (\ref{pw2}) is not well-defined, so we need to fix the initial condition in this case in a different way. It can be shown from Eq.(\ref{fe}) that $a^{\prime\prime}/a$ is a constant if $\rho\propto 1/a^2$. Therefore, we can define a new variable  $k^\prime\equiv\sqrt{k^2-a^{\prime\prime}/a}$ and interpret $k^\prime$ as an effective comoving wave number.  The wave equation (\ref{waveeq}) then becomes
\begin{equation}
\frac{d^2u_k}{d\tau ^2}+k^{\prime2}u_k=0,
\end{equation}
which has two properly normalized linearly independent solutions,
\begin{equation}
u_k(\tau,k)=\frac{e^{-ik^\prime\tau}}{\sqrt{2k^\prime}}, \,\,\,\,  \frac{e^{ik^\prime\tau}}{\sqrt{2k^\prime}}.
\label{csini}
\end{equation}
We choose the exponent with the minus sign because it reduces to the Minkowski initial condition \cite{Langlois:2010xc}.
\end{itemize}

Finally, it is easier to use the scale factor as the independent variable in the numerical integration of Eq.~(\ref{wavesf}) instead of  the conformal time. The relation between the conformal time $\tau$ and the scale factor $a$ is
\begin{eqnarray}
\tau& =&\frac{\sqrt{3}}{\kappa }\frac{1}{h\left(\beta _1-\zeta \right)}B_1{}^{-1/2}\left(\frac{B_1}{A_2{}^{1\left/\left(1+\beta _3\right)\right.}}\right)^hx^h
\nonumber\\
&\,&_2F_1\left[h, 1-g, h+1; x\right],
\end{eqnarray}
where $x=1-[1+(A_2^{1/(1+\beta _3)}/B_1)a^{\beta _1-\zeta}]^{-1}$, $g=(1-1/(2\zeta ))/(\beta _1-\zeta )$, and $e=(\beta _1/2-1)/(\beta _1-\zeta )$.

\begin{figure}[t]
\centering
\includegraphics[width=8cm]{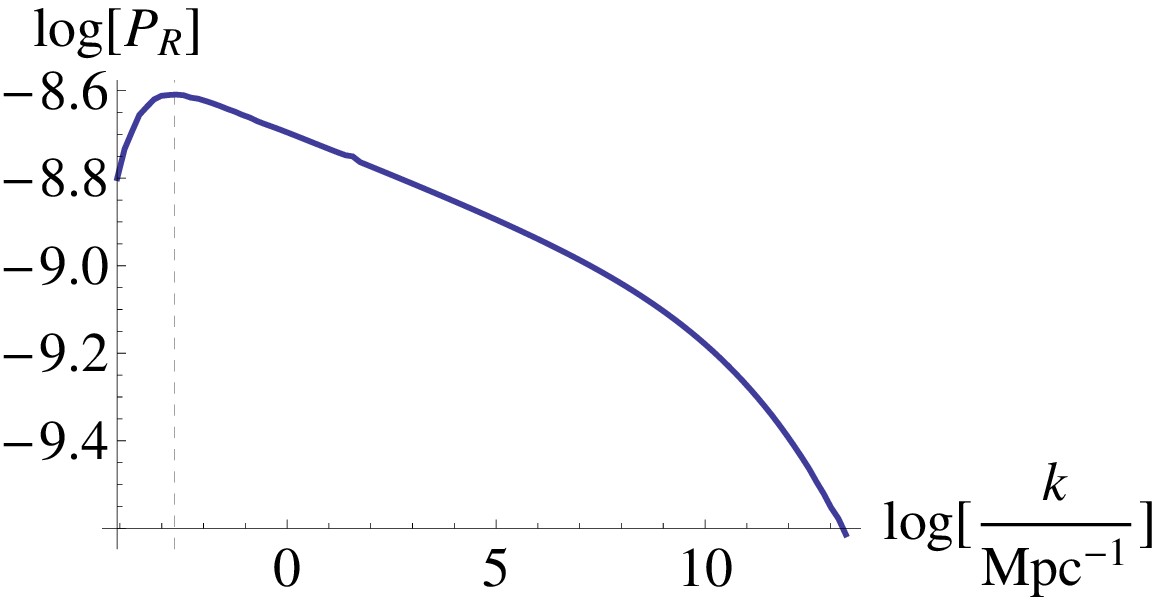}
\centering
\caption{This plot corresponds to the curvature perturbation spectrum for
 the inspired modified GCG model with $\beta_1=1$ (see Eq.~(\ref{2m1})), which describes a NFDW dominated era followed by a power-law inflationary period. We choose $\beta_3=-1.05$. The vertical dashed line corresponds to the pivot scale $k=0.002$ Mpc$^{-1}$.}
\label{DWS}
\end{figure}

\begin{figure}[t]
\centering
\includegraphics[width=8cm]{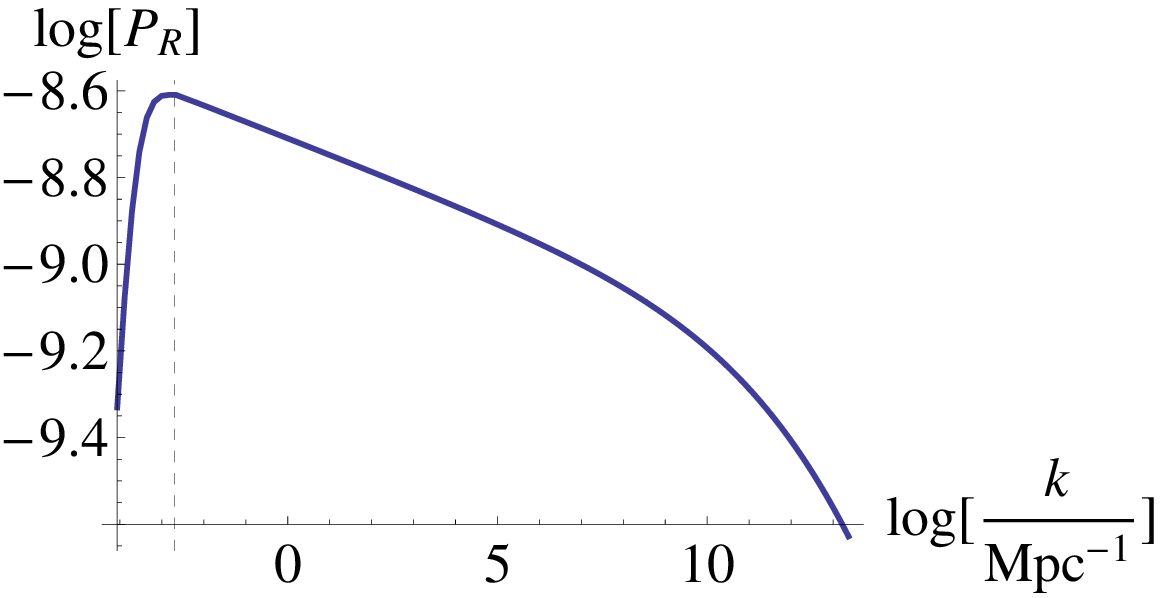}
\centering
\caption{This plot corresponds to the curvature perturbation spectrum for the modified GCG model with $\beta_1=2$ (see Eq.~(\ref{2m1})), which describes a NFCS dominated era followed by a power-law inflationary period. We choose $\beta_3=-1.05$. The vertical dashed line corresponds to the pivot scale $k=0.002$ Mpc$^{-1}$.}
\label{CSS}
\end{figure}

The resulting curvature power spectra are shown in Fig.~\ref{DWS} and Fig.~\ref{CSS}, corresponding to the NFDW and the NFCS, respectively. Let us recall that all the parameters used here have been fixed by imposing the observational constraints in the way  stated in Sec.~\ref{model}. For modes $k$ such that $10^{-3}\,\textrm{Mpc}^{-1}<k<10^{5}\,\textrm{Mpc}^{-1}$, we obtain a constant slope on the logarithmic amplitude of the curvature perturbation. This is a simple consequence of the intermediate phase given in Eq.(\ref{2m2}) previously analyzed in Ref.~\cite{BouhmadiLopez:2011kw}, and implies a power-law inflation. Our new and important result is the drop of $P_R$ for the modes whose $k\leq$10$^{-3}$Mpc$^{-1}$, which is helpful in explaining the quadrupole anomaly through an alternative way from those used in Refs.~\cite{Contaldi:2003zv,Boyanovsky:2006pm,Powell:2006yg,Scardigli:2010gm,Piao:2005ag}. Such a decrease of $P_R$ is a consequence of the NFTD era just before the inflation.


We calculate the CMB temperature anisotropy spectrum by the numerical package CMBFAST \cite{Seljak1996, Zaldarriaga1998, Zaldarriaga2000} with minor modifications to the form of specifying the primordial power spectrum. In the modified version of CMBFAST, the primordial power spectrum is fed into the code as an interpolating function instead of a functional form. This adjustment is made so that rather than accepting only the nearly-scale-invariant spectrum as its initial condition, CMBFAST is now compatible with any general shape of initial spectrum. This feature is essential to our case since the primordial spectra obtained from our scenarios severely deviate from the scale-invariant form in large scales whose wave numbers are smaller than $10^{-3} \, \textrm{Mpc}^{-1}$.

\begin{figure}

\centering

\includegraphics[width=8cm]{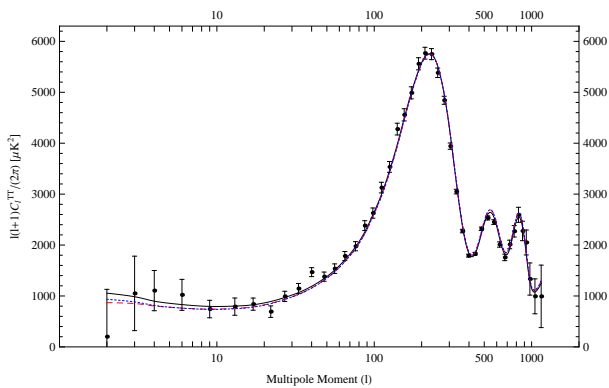}

\caption{
	The CMB temperature anisotropy spectrum. The dots with error bars are the WMAP 7-year data. The solid line is the prediction of standard inflation with a power law spectrum. The dashed and dotted lines are the spectra of the scenarios of NFCS and NFDW, respectively.
}

\label{fig:CMBSpectrum}

\end{figure}

Fig.~\ref{fig:CMBSpectrum} shows the CMB spectra generated by the NFTD scenarios along with the standard inflationary model. The data of WMAP 7-year observation \cite{Komatsu:2010fb} are also marked in the plot. It can be seen that the pre-inflation NFTD era alleviates the quadrupole anomaly of the CMB. Note that the NFCS has a stronger effect on reducing the amplitude of the lower modes than the NFDW does. This is a result of the initial slope and the turn-around point of the curvature perturbation spectrum induced by NFDW and NFCS as shown in Fig.~\ref{DWS} and \ref{CSS}. Also note that regarding the lower modes of the CMB, it is irrelevant that the potentials and the first derivatives of the scalar field with respect to the cosmic time are not rigorously continuous at the connecting point. The reason  is that major contributions to the lower modes of CMB came from the scalar perturbations whose comoving wave numbers are about $10^{-5}$ to $10^{-3} \, \text{Mpc}^{-1}$. {\color{black}T}hese modes had already exited the Hubble radius {\color{black}during} the first period so that the power spectrum in this regime is obtained without {\color{black}the need to integrate} across the connecting point.




\section{Tensor perturbation}

\label{tp}
As it is well known within the inflationary paradigm, aside from the scalar perturbations, there are also tensorial perturbations which give rise to a stochastic background of gravitational waves generated from the vacuum quantum fluctuations during the inflationary era.  If the stochastic background of GWs of cosmological origin is ever detected (hopefully on the next decades \cite{Smith:2005mm, Kawasaki:2012rw}), it will provide an amazing amount of information and additional probes about the very early universe. We next obtain the GWs power spectra predicted by the model described in Eq.~(\ref{2m}). Our analysis will be based on the method of the Bogoliubov coefficients and this study is motivated by the fact that we hope the detection of the GWs power spectrum can throw some light on the physics of the early universe, as we have already mentioned.

\subsection{Formalism}
The gravitational waves energy density is given by \cite{Sa:2007pc}
\begin{equation}
\rho _{\text{GW}}=\int \left\langle N_k(\tau )\right\rangle (2\hbar \omega )\frac{\omega ^2}{2\pi ^2c^3}d\omega,
\end{equation}
where $ \left\langle N_k(\tau )\right\rangle$ is the number of gravitons created for a given mode $k$ in a different vacuum state as the universe expands. Then the dimensionless relative logarithmic energy spectrum $\Omega_{\text{GW}}$ at the present time $\tau_0$ can be defined as \cite{Sa:2007pc}
\begin{eqnarray}
\Omega_{\text{GW}}(\omega,\tau_0)\equiv\frac{1}{\rho_c(\tau_0)}\frac{d\rho_{\text{GW}}}{d\ln\omega}(\tau_0)\nonumber\\=\frac{\hbar\kappa^2}{3\pi^2 c^5 H^2(\tau_0)}\omega^4 \left\langle N_k(\tau_0 )\right\rangle^2.
\label{spectrum}
\end{eqnarray}
We will use the method of the Bogoliubov coefficients to obtain the number of gravitons created as the universe expands  \cite{Moorhouse:1994nc}.

The tensor perturbation $h_{ij}$ can be decomposed into Fourier modes $\xi_k$ which obey:
\begin{equation}
\frac{d^2\xi_k}{d\tau ^2}+\left(k^2-\frac{1}{a}\frac{d^2a}{d\tau ^2}\right)\xi_k=0.
\label{gweq}
\end{equation}
The initial vacuum state changes as the universe expands. This change can be described through a Bogoliubov transformation:
\begin{equation}
a(k)=\alpha_k A(k)+\beta_k A^\dag(-k),
\label{svacuum}
\end{equation}
where $\alpha_k$ and $\beta_k$ are the Bogoliubov coefficients. In particular, the  coefficient $\beta_k$ gives the number of gravitons created for each mode $k$; i.e. $\left\langle N_k(\tau )\right\rangle=|\beta_k(\tau)|$. Given that the universe was initially in a vacuum state, we  impose the initial conditions that $\alpha_k=1$ and $\beta_k=0$, which corresponds to the absence of gravitons at the beginning. In addition, $A$ and $A^\dag$ correspond to the annihilation and creation operators of the gravitons.

\subsection{Numerical Solutions}
The evolution equation for the Bogoliubov coefficients can be simplified through the substitutions \cite{Moorhouse:1994nc}
\begin{eqnarray}
\alpha_k(\tau)=\frac{X(k;\tau )+Y(k;\tau )}{2\xi_{k}(\tau )}, \label{alphak}  \\
\beta_k(\tau)=\frac{X(k;\tau )-Y(k;\tau )}{2\xi_{k}^*(\tau )},
\label{betak}
\end{eqnarray}
where $\xi_k$ is the appropriate boundary condition chosen for a given mode $k$. A discussion parallel to that given in Sec.~\ref{sp} follows:

\begin{itemize}

\item For the modes with comoving wave numbers roughly smaller than $10^{-3}$ Mpc$^{-1}$, the integration starts during the NFTD dominated era where the condition $k\gg a H$ is satisfied and where the mode is well inside the horizon. Therefore, we can just use the Minkowski initial condition \cite{Langlois:2010xc}
\begin{equation}
 \xi_k\rightarrow\frac{e^{ik\tau}}{\sqrt{2k}}
\label{mkini}
\end{equation}
for the NFDW period. For the NFCS, we use the initial condition given in Eq.~(\ref{csini}).

\item For the modes with comoving wave number roughly larger than $10^{-3}$ Mpc$^{-1}$, we can start the integration during the power-law inflationary dominated era where the condition $k\gg a H$ is satisfied and where the mode is well inside the horizon. Thus we can also use the Minkowski initial condition in Eq.~(\ref{mkini}).

\end{itemize}
\begin{figure}[t]
\centering
\includegraphics[width=8cm]{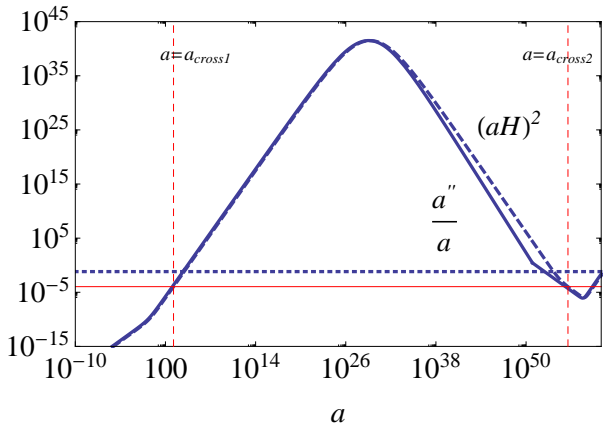}
\centering
\caption{The integration for a given mode $k$. The dashed curve corresponds to $(a H)^2$ and the solid curve corresponds to $a''/a$ for a modified $\beta_1=1$ GCG model. The horizontal red line shown in the figure corresponds to the pivot scale $k=0.002$ Mpc$^{-1}$, and the horizontal dashed blue line corresponds to the intersection of the first two periods. }
\label{DW}
\end{figure}
\begin{figure}[t]
\centering
\includegraphics[width=8cm]{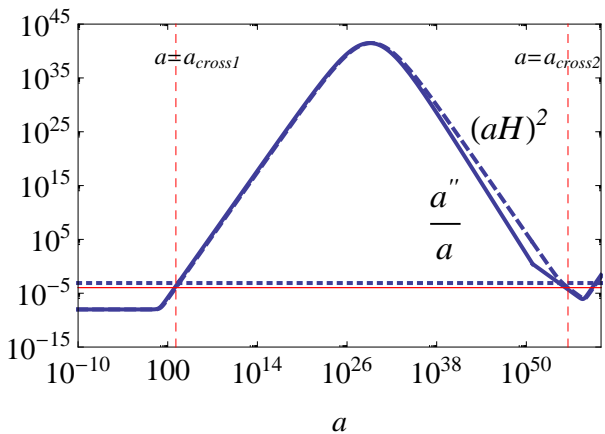}
\centering
\caption{The integration for a given mode $k$. The dashed curve corresponds to $(a H)^2$ and the solid curve corresponds to $a''/a$ for a modified $\beta_1=2$ GCG model. The horizontal red line shown in the figure corresponds to the pivot scale $k=0.002$ Mpc$^{-1}$, and the horizontal dashed blue line corresponds to the intersection of the first two periods. }
\label{CS}
\end{figure}
By substituting Eqs.~(\ref{alphak})-(\ref{betak}) into Eq.~(\ref{gweq}) and using the initial conditions of $\xi_k$ chosen above, Eq.~(\ref{gweq}) can be simplified to a set of first order linear differential equations of the variables $X$ and $Y$ \cite{Moorhouse:1994nc}
\begin{eqnarray}
\begin{cases}
 Y=\frac{i}{k}X', \\
 X^{\prime\prime }+\left(k^2-\frac{a^{\prime\prime }}{a}\right)X=0.
\end{cases}
\label{simplified}
\end{eqnarray}
After integrating Eq.~(\ref{simplified}), we use Eq.~(\ref{betak}) to obtain the number of gravitons created till the present time $\tau_0$
\begin{equation}
\left\langle N_k(\tau )\right\rangle=|\beta_k(\tau_0)|=\left.\frac{|X-Y|}{2}\frac{1}{\sqrt{\xi_{\nu }^*\xi_{\nu }}}\right.|_{\tau=\tau_0}.
\end{equation}

\begin{figure}[t]
\centering
\includegraphics[width=8cm]{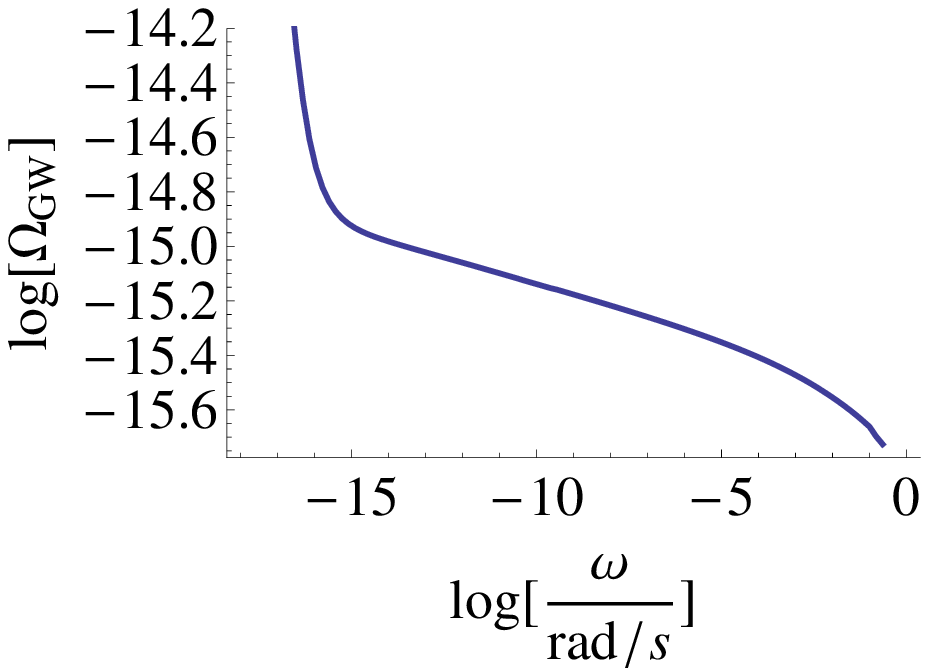}
\centering
\caption{This plot corresponds to the  gravitational wave spectrum for
 the inspired modified GCG model with $\beta_1=1$ (see Eq.~(\ref{2m1})), which describes a NFDW dominated era followed by a power-law inflationary period with $\beta_3=-1.05$.}
\label{DWT}
\end{figure}

\begin{figure}[t]
\centering
\includegraphics[width=8cm]{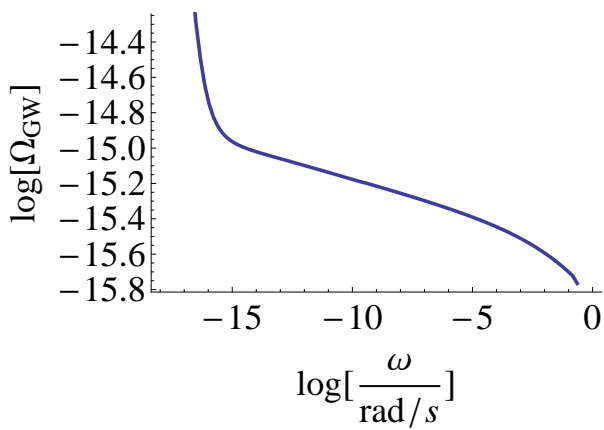}
\centering
\caption{This plot corresponds to the  gravitational wave spectrum for
 the inspired modified GCG model with $\beta_1=2$ (see Eq.~(\ref{2m1})), which describes a NFCS dominated era followed by a power-law inflationary period with $\beta_3=-1.05$.}
\label{CST}
\end{figure}

The integration method can be read from Fig. \ref{DW} and Fig. \ref{CS} for NFDW and NFCS, respectively. For a given mode $k$, we start the integration at the time when the scale factor $a$ is a thousandth of the scale where the mode exits the horizon; i.e. $a_{\rm start}=0.001a_{\rm cross_1}.$ The integration ends at the time when the scale factor $a$ is thousand times of the scale where the mode reenters the horizon; i.e. $a_{\rm end}=1000a_{\rm cross_2}$. The reason for stopping our numerical calculations after the modes are well inside the horizon and not continuing the numerics till the present time is the following: once a mode is inside the horizon, the mode begins to oscillate and the number of gravitons created will not change anymore. Therefore, we can save a lot of computational time with this methodology \cite{BouhmadiLopez:2009hv}. The resulting spectra are shown in Fig. \ref{DWT} and Fig. \ref{CST}. The frequencies range approximately from $10^{-17}$ to $10^0$ Hz, and the dimensionless relative logarithmic energy spectrum of the GWs falls approximately between $10^{-14}$ and $10^{-16}$. This result  fulfills the current observational upper limits (all solid curves for various gravitational-wave detectors) shown in Fig.~2 of Ref.~\cite{Smith:2005mm}. The gravitational waves predicted in our model could be detected in future experiments by some planned detectors like BBO and DECIGO whose sensitivities are shown in Fig.~6 of Ref.~\cite{Kawasaki:2012rw}.
It was shown in Ref.~\cite{BouhmadiLopez:2011kw} (see also Ref.~\cite{BouhmadiLopez:2009hv}) that only large frequencies are sensitive to the slow-roll regime controlled by the parameter $\beta_3$. For those frequencies, a larger $\beta_3$  gives rise to a higher maximum of the potential $a''/a$, a fact that translates into a shift of the power spectrum of the GWs towards the right.
For middle frequencies, the plateau of the spectrum merges for different $\beta_3$ since the energy scale of  inflation is almost fixed and independent of the values acquired by $\beta_3$. Similarly, we have shown that for low frequencies, the spectrum merges since the energy scale of the NFTD is almost indifferent to the changes of the parameter $\beta_3$.

\section{Summary}
\label{sum}

The goal of this paper is to propose a possible solution to the low quadruple of the CMB which is based on a NFTD period just before the slow-roll inflationary era. The NFTD can be of two kinds: (i) a NFDW implying a slower inflationary rate than that implied by the standard slow-roll inflation, or (ii) a NFCS describing a universe sitting on the border line between acceleration and deceleration.
We model such a  matter content for the very early universe as a kind of a generalized Chaplygin gas described by Eq.~(\ref{2m}) (see also Eq.~(\ref{1.1})), which gives a smooth transition from a NFTD dominated era to a de Sitter-like phase or a power-law inflationary era. We constrain our model using the  WMAP7 data for the power spectrum of the scalar perturbations, $P_s=2.45\times10^{-9}$, and the spectral index, $n_s=0.963$, at the pivot scale, $k_0=0.002$ Mpc$^{-1}$ \cite{Komatsu:2010fb}. After fixing the parameters of the model and imposing suitable boundary conditions for the perturbations, we obtain the curvature power spectra for the cases of a NFDW and a NFCS, as shown in Figs.~\ref{DWS} and \ref{CSS}. The most important feature of the spectrum is the drop of the power starting from $k_0\sim 0.002 $ Mpc$^{-1}$. The reason behind this drop is the high energy effects of the NFTD on the lowest modes. Therefore, this mechanism can alleviate the observed low CMB quadrupole mode \cite{Contaldi:2003zv,Boyanovsky:2006pm,Powell:2006yg,Wang:2007ws,Scardigli:2010gm}, as shown in Fig.~\ref{fig:CMBSpectrum}. Besides, we obtain the power spectra of the gravitational wave predicted by our model, which are consistent with the tensor-to-scalar ratio of the current observation \cite{Smith:2005mm,Kawasaki:2012rw}. The regime of frequencies roughly larger than $10^{-5}$ Hz is within the reach of planned detectors such as BBO and DECIGO. Those frequencies are sensitive to the only degree of freedom of our model described by the parameter $\beta_3$. As a result, $\beta_3$ could be constrained by the projected sensitivities of BBO and DECIGO.

In summary, even though the model presented here could be seen as an over simplified toy model, it can give a guiding light towards a more consistent picture of a topological defect era prior to the slow-roll inflationary one, which could alleviate the observed CMB anomalies such as the suppression of the $l=2$ quadruple mode of the CMB.

\acknowledgments

The work of M.B.L. was supported by the Spanish Agency ``Consejo Superior de Investigaciones Cient\'{\i}ficas" through JAEDOC064 and the Basque Foundation for Science IKERBASQUE. She also wishes to acknowledge the hospitality of LeCosPA Center at the National Taiwan University during the completion of part of this work and the support of the Portuguese Agency ``Funda\c{c}\~{a}o para a Ci\^{e}ncia e Tecnologia" through PTDC/FIS/111032/2009.
P.C.,  Y.C.H. and Y. H. L.  are supported by Taiwan National Science Council under Project No. NSC 97-2112-M-002-026-MY3 and by Taiwan’s National Center for Theoretical Sciences (NCTS). P.C. is in addition supported by US Department of Energy under Contract No. DE-AC03-76SF00515. This work has been supported by a Spanish-Taiwanese Interchange Program with reference 2011TW0010 (Spain) and NSC 101-2923-M-002-006-MY3 (Taiwan).

\end{document}